\documentclass[doublecol,final]{epl2} 
\usepackage{graphicx}
\usepackage{subcaption}
\usepackage{amsmath,amssymb,amsfonts}
\usepackage{sidecap}
\usepackage{xcolor}
\usepackage{lipsum}
\usepackage{physics}
\usepackage{hyperref}

\newcommand{\Var}{\mathrm{Var}}

\title{Universal thermodynamic bounds on the Fano factor of discriminatory networks with unidirectional transitions}
\shorttitle{Universal thermodynamic bounds on the Fano factor of discriminatory networks} %Insert here a short version of the title if it exceeds 70 characters

\author{J. Berx\thanks{E-mail: \email{jonas.berx@nbi.ku.dk}} \and K. Proesmans}
\shortauthor{J. Berx \etal}

\institute{                    
  Niels Bohr International Academy, Niels Bohr Institute, University of Copenhagen, Blegdamsvej 17, 2100 Copenhagen, Denmark
}

\abstract{
We derive a universal lower bound on the Fano factors of general biochemical discriminatory networks involving irreversible catalysis steps, based on the thermodynamic uncertainty relation, and compare it to a numerically exact Pareto optimal front. This bound is completely general, involving only the reversible entropy production per product formed and the error fraction of the system. We then show that by judiciously choosing which transitions to include in the reversible entropy production, one can derive a family of bounds that can be fine-tuned to include physical observables at hand. Lastly, we test our bound by considering three discriminatory schemes: a multi-stage Michaelis-Menten network, a Michaelis-Menten network with correlations between subsequent products, and a multi-stage kinetic proofreading network, where for the latter application the bound is altered to include the hydrolytic cost of the proofreading steps. We find that our bound is remarkably tight.
}

\begin{document}
\graphicspath{ {Images/} }

\maketitle
\section{Introduction}\label{sec:intro}

Discriminatory processes, such as proofreading mechanisms, that select particular molecules from a plethora of others are of seminal importance in structural biology, and form the basis of many of the information processing tools fundamental to complex life \cite{Alon2006}. The need for high fidelity in biological systems is intimately linked to non-equilibrium stochastic processes, where affinities drive chemical reactions away from equilibrium. 

These enzymatic processes are generally subjected to strong thermal fluctuations. To characterise the fluctuations of product molecules resulting from these reactions, one can use a measure of the dispersion of a counting process, known as the Fano factor (alternatively, the randomness parameter or coefficient of variation in the long-time limit) \cite{CHOWDHURY20131,Moffitt2014,Proesmans2019}.

Single-molecule experiments in biophysical systems  have made it possible to obtain precise measurement data on the dynamics of single enzymes \cite{Ritort2006,Cornish2007}. In the context of discrimination, the Fano factor has for instance been used to characterise fluctuations of the timing of SOS activation \cite{Huang2019,Huang2021}, establishing a basis for kinetic proofreading (KPR) \cite{Hopfield1974,Ninio1975,Song2021} in the receptor-mediated activation of Ras guanine nucleotide exchange factors (GEFs), and in a model for rapid DNA scanning \cite{Ray_2021}, followed by `snap-locking' in Holliday junctions formed during genetic recombination, which is consistent with conformational proofreading \cite{Savir_2007}. 

As such, the precise relation between the Fano factor and other kinetic or thermodynamic quantities such as the error or the entropy production rate becomes of the utmost importance in the characterisation and optimal design of discriminatory networks. In this work, we aim to find general bounds on the Fano factor involving the aforementioned variables, based on the thermodynamic uncertainty relation (TUR) \cite{Barato2015,Barato2015_2,gingrich2016dissipation,horowitz2020thermodynamic}. This relation states that there is a fundamental trade-off between the uncertainty (or fluctuations) of a stochastic variable $X$, and the entropy production. 

The original TUR has been applied to a number of biochemical processes \cite{Proesmans2019,hwang2018energetic,kim2021thermodynamic} including single-stage kinetic proofreading (KPR) schemes \cite{Pineros2020}, where all transition are assumed to be fully reversible.

However, many enzymatic reaction networks, including discriminatory systems are modelled with explicit unidirectional steps, as the reverse process rarely occurs \cite{Rao_2015}. Although this simplifies the model, it makes it impossible to apply the original TUR. Even if one includes an explicit reversed rate, the TUR is generally expected to be very loose \cite{horowitz2020thermodynamic}.
%The extension of the Fano factor TUR to systems involving irreversible transitions is pertinent in \emph{e.g.}, statistical kinetics, where such transitions are modelled using the formal limit of diverging thermodynamic affinity $\mathcal{A}$, a reasonable assumption for experiments performed at high affinity. 

We will therefore use an alternative TUR that allows for uni-directional transitions \cite{Pal2021}. This allows us to infer a simple and general trade-off relation between the Fano factor of the production molecule, the error rate and the entropy production.
We test our results on three distinct discriminatory systems:  a multi-stage Michaelis-Menten network, a Michaelis-Menten network with correlations between subsequent products, and a multi-stage kinetic proofreading network. We check our bound's tightness by means of a numerical Pareto optimal front analysis \cite{Chiuchiu_2023}.

 %Our study can be viewed as complementary, and when taking the limit of irreversible catalysis in a single stage network, their results can be generalised to ours.

\section{Thermodynamic uncertainty relations}\label{sec:TUR}

\begin{figure}[tp]
    \centering
    \includegraphics[width=0.8\linewidth]{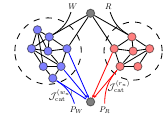}
    \caption{General discriminatory network where an enzyme $E$ (gray vertices) can bind either correct $R$ (red vertices) or wrong $W$ (blue vertices) substrates to yield the corresponding products $P_R$ and $P_W$. The network need not be symmetric under interchange $R\leftrightarrow W$, and the number of catalysis steps may vary. The unbound enzymatic states are identified.}
    \label{fig:universal_network}
\end{figure}

Let us set the stage by considering a general discriminatory network that converts substrate molecules $S$ into product molecules $P_S$, with $S\in\{R,\,W\}$ denoting the correct $(R)$ or wrong $(W)$ substrate or product. A simple cartoon is sketched in fig.~\ref{fig:universal_network}. Note that the free enzymatic states, denoted by gray vertices, pertain to the same state $E$.

To proceed, we only require that the catalysis steps leading to the formation of product $P_S$ are completely irreversible.

Such discriminatory networks are usually supplemented with the following assumptions, none of which are crucial for our general results to hold: i) the networks are generally assumed to be symmetric with respect to the interchange $R\leftrightarrow W$, \emph{i.e.}, the discrimination takes place on the level of kinetic rates, not on the level of the network structure; ii) the catalysis transition rates are assumed to be equal for the different production steps, and iii) only one catalysis transition can lead to every distinct product. We will later show an example that breaks the latter assumption.

We study steady-state fluctuations in the number of product molecules that are being formed by means of the TUR. In steady state at a temperature $T$, the TUR, as it was formulated originally \cite{Barato2015} is given by
\begin{equation}
    \label{eq:TUR_bi}
    \mathcal{Q}\equiv T\sigma \tau \epsilon^2 \geq 2 k_B T
\end{equation}
where $\epsilon^2 \equiv (\langle X^2\rangle -\langle X\rangle^2)/\langle X\rangle^2 = \Var(X)/\langle X\rangle^2$ is the squared relative uncertainty of the stochastic generalised current $X$ and $\sigma$ is the entropy production rate, in units of $k_{\rm B}$. The total dissipation after a time $\tau$ is then $T\sigma\tau$. Henceforth, we will assume all energies are measured in units of $k_{\rm B}T$, so we will fix $k_{\rm B} = T=1$.

Since the form \eqref{eq:TUR_bi} requires that all transitions are reversible and obey local detailed balance, it is not suitable for processes where one or more transitions are unidirectional, since these would lead to an infinite entropy production. Recently, however, the TUR has been extended to accommodate such systems, by means of the Cramér-Rao inequality \cite{Dechant2018,Hasegawa2019,Ito2020} and explicitly taking the various irreversible steps into account \cite{Pal2021}. For a fluctuating current $X$ in a time window of length $\tau$, with mean flux $\mathcal{J}(\tau) = \langle X\rangle/\tau$, the generalised TUR is given by
\begin{equation}
    \label{eq:TUR_uni}
    \Var(X) \geq \frac{\tau^2 \mathcal{J}^2(\tau)}{\int_0^\tau\mathrm{d}t [\frac{\sigma_{\rm rev}(t)}{2}+\mathcal{J}_{\rm uni}(t)]}\,,
\end{equation}
where $\sigma_{\rm rev}(t)$ is the entropy production rate of the reversible transitions, and $\mathcal{J}_{\rm uni}(t)$ is the average current of all unidirectional transitions. If no unidirectional currents are present, the TUR reduces to the original \eqref{eq:TUR_bi}. In the steady state, the entropy production and currents are time-independent, and the generalised TUR can be rewritten as follows
\begin{equation}
    \label{eq:TUR_uni_ss}
    \Var(X) \geq \frac{\tau \mathcal{J}^2}{\frac{\sigma_{\rm rev}}{2}+\mathcal{J}_{\rm uni}}\,.
\end{equation}
The diffusion coefficient $D$ of a stochastic variable in the steady state, observed for a time $\tau$, is defined by $\Var{(X)} = 2\tau D$. Inserting this into eq.~\eqref{eq:TUR_uni_ss}, the generalised TUR now takes the final form
\begin{equation}
    \label{eq:TUR_uni_final}
    \mathcal{F} \equiv \frac{2 D}{\mathcal{J}} \geq \frac{\mathcal{J}}{\frac{\sigma_{\rm rev}}{2}+\mathcal{J}_{\rm uni}}\,,
\end{equation}
where $\mathcal{F}$ is the Fano factor corresponding to $X$.

To proceed, we define a notion of the error fraction $\eta$ for general processes involving different catalysis transitions with associated unidirectional fluxes $\mathcal{J}_{\rm cat}^{(\gamma)}$ leading to either correct or wrong substrates, where we define the former as the set where $\gamma\in\mathcal{R} = \{r_1,r_2,...,r_n\}$ and the latter as $\gamma\in\mathcal{W} = \{w_1,w_2,...,w_m\}$, with $m$ not necessarily equal to $n$. The error fraction is the ratio of catalysis transitions leading to the wrong products and the total number of catalysis transitions, \emph{i.e.},
\begin{equation}
    \label{eq:error_general}
    \eta = \frac{\sum\limits_{\gamma\in\mathcal{W}}\mathcal{J}_{\rm cat}^{(\gamma)}}{\sum\limits_{\gamma\in\mathcal{W}\cup\mathcal{R}}\mathcal{J}_{\rm cat}^{(\gamma)}}\,.
\end{equation}

If we now want to study the fluctuations in the number of products formed, we need to count the total number of times the catalysis transitions leading to these products occur, \emph{i.e.}, we will study the variance of the random variables $X_R,\,X_W$, defined as $X_R \equiv \sum\limits_{\gamma\in\mathcal{R}}X_\gamma$ and $X_W \equiv \sum\limits_{\gamma\in\mathcal{W}}X_\gamma$. Assuming that all other transitions in the network are formally reversible and obey local detailed balance, the total unidirectional flux can be written as 
\begin{equation}
    \label{eq:unidirectional_fluxes}
    \mathcal{J}_{\rm uni} = \mathcal{J}_R + \mathcal{J}_W = \sum\limits_{\gamma\in\mathcal{R}}\mathcal{J}_{\rm cat}^{(\gamma)} + \sum\limits_{\gamma\in\mathcal{W}}\mathcal{J}_{\rm cat}^{(\gamma)}\,.
\end{equation}
Plugging eq.~\eqref{eq:unidirectional_fluxes} into \eqref{eq:TUR_uni_final} yields the following bound on the Fano factors for the catalysis transitions,
\begin{equation}
\begin{split}
    \mathcal{F}_S &\geq \frac{\sum\limits_{\gamma\in\mathcal{S}}\mathcal{J}_{\rm cat}^{(\gamma)}}{\frac{\sigma_{\rm rev}}{2}+\sum\limits_{\gamma\in\mathcal{R}}\mathcal{J}_{\rm cat}^{(\gamma)}+\sum\limits_{\gamma\in\mathcal{W}}\mathcal{J}_{\rm cat}^{(\gamma)}}\,,
\end{split}
\end{equation}
where $S\in\{R,W\}$ and $\mathcal{S}\subset\{\mathcal{R},\mathcal{W}\}$, correspondingly. If we now define the reversible entropy production per product formed as $\Delta\sigma = \sigma_{\rm rev}/(\mathcal{J}_R + \mathcal{J}_W)$, the bounds on the Fano factors for correct and wrong product steps can be concisely written as
\begin{equation}
    \label{eq:TUR_Fano}
    \mathcal{F}_R \geq \frac{1-\eta}{1+\frac{\Delta\sigma}{2}}\,,\qquad \mathcal{F}_W \geq \frac{\eta}{1+\frac{\Delta\sigma}{2}}\,,
\end{equation}
respectively, which constitutes our main result. It can be seen that increasing the error fraction to unity pushes the lower bound for the correct product Fano factor to zero, allowing for high predictability, since no correct products are formed. Similarly, lowering the error fraction to zero allows for high predictability of the number of wrong products. For $\eta\downarrow0$, however, the entropy production diverges to infinity for general discriminatory networks \cite{Rao_2015}, and the lower bound tends to zero once again, showing that in the limit of small error rates, high predictability of the number of correct products formed is necessarily coupled to increasing entropy production. 

Note that the bound \eqref{eq:TUR_Fano} is completely general for all types of discrimination processes where the only unidirectional transitions are the production steps for right and wrong products. The topology of the reversible part of the network is irrelevant for the bound to hold. 

From the above analysis, we can moreover derive the following corollary. The fluctuations of the number of product molecules, without distinguishing between $R$ or $W$, \emph{i.e.}, of the stochastic variable $X_P \equiv X_R + X_W$, are bounded from below by \eqref{eq:TUR_uni_ss},
\begin{equation}
    \label{eq:TUR_combined}
    \Var{(X_P)} \geq \frac{\tau (\mathcal{J}_R + \mathcal{J}_W)^2}{\frac{\sigma_{\rm rev}}{2}+\mathcal{J}_{\rm uni}}\,.
\end{equation}
Consequently, the entropy production per product is bounded from below by 
\begin{equation}
    \label{eq:TUR_combined_dissipation}
    \Delta\sigma \geq 2 \left(\frac{1}{\mathcal{F}_P}-1\right)\,,
\end{equation}
with $\mathcal{F}_P \equiv 2 D_{R+W}/(\mathcal{J}_R + \mathcal{J}_W)$. This bound only yields a positive lower bound for $\mathcal{F}_p<1$, \emph{i.e.}, for a peaked production cycle time distribution. For comparison, the bound reduces to $\Delta\sigma\geq 2/\mathcal{F}_P$ for completely reversible discrimination processes.

We now turn to some simple applications to illustrate our bounds for different biochemical network topologies. In all the applications, expressions for the production fluxes and diffusion coefficients are obtained analytically by means of Koza's steady-state method \cite{Koza1999}, and the reversible entropy production \cite{Rao_2015},
\begin{equation}
    \label{eq:entropy_production_general}
    \sigma_{\rm rev} = \frac{1}{2}\sum\limits_{i,j}^{}\,' (k_{ij} P_i^s - k_{ji} P_j^s) \ln\frac{k_{ij} P_i^s}{k_{ji} P_j^s}
\end{equation}
is calculated by solving the corresponding master equation 
\begin{equation}
    \label{eq:master_equation_general}
    \dv[]{\mathbb{P}}{t} = {\bf K}\,\mathbb{P}\,,
\end{equation}
with transition matrix ${\bf K}$ exactly for the steady-state probability distributions $P_i^s$. In eq.~\eqref{eq:entropy_production_general}, $k_{ij}$ is the transition rate from $i\rightarrow j$, and the primed sum runs over all pairs of states $i,\,j$ but excludes the irreversible (catalysis) transitions. 

We supplement our results with a Pareto optimal front analysis of the discriminatory systems, based on a genetic algorithm that performs simultaneous multi-function optimisation on objective functions, by varying the kinetic rates \cite{Deb2001}. The resulting Pareto optimal front $\mathcal{P}$ illustrates the optimal trade-off between the Fano factor and the quantity on the r.h.s.~of eq.~\eqref{eq:TUR_Fano}, which we will henceforth call $\zeta$, where further optimisation of one objective is to the detriment of the other. 

\section{Biochemical applications}\label{sec:examples}
\subsection{Multi-stage Michaelis-Menten (MM) network}\label{subsec:MM}
\begin{figure}[tp]
    \centering
    \includegraphics[width=\linewidth]{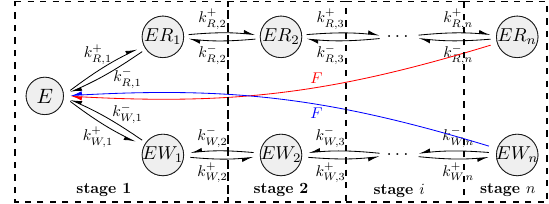}
    \caption{The multi-stage MM discriminatory network with $n$ conformational bound states per substrate. The irreversible catalysis steps are shown in red (blue) for correct (wrong) product formation.}
    \label{fig:MM_network}
\end{figure}

The simplest application we will consider to test our bounds is the $n-$stage MM reaction scheme, illustrated in fig.~\ref{fig:MM_network}, where the enzyme $E$ can bind a substrate $S\in\{R,\,W\}$ to form the complex $ES_1$. This complex can then reversibly go through multiple conformational stages $i=1,2,...,n$, until it irreversibly forms product $P_S$ from the state $ES_n$ with catalytic rate $F$, which we assume to be identical for both correct and wrong substrates.
In this scheme, the error rate $\eta$ is given by
\begin{equation}
    \label{eq:MM_error}
    \eta = \frac{\mathcal{J}_W}{\mathcal{J}_R+\mathcal{J}_W} = \frac{P_{ER_n}^s}{P_{ER_n}^s+P_{EW_n}^s}\,,
\end{equation}
where $\mathcal{J}_R,\,\mathcal{J}_W$ are the correct and wrong product fluxes, respectively. These production fluxes are the only irreversible fluxes in the system, such that $\mathcal{J}_{\rm uni} = \mathcal{J}_R + \mathcal{J}_W$. $P_{ER_n}^s,\,P_{EW_n}^s$ are the steady-state probability distributions for the enzyme to be in the $ER_n,\,EW_n$ bound conformational states. The reversible entropy production $\sigma_{\rm rev}$ is given by
\begin{equation}
    \label{eq:MM_entropy}
    \begin{split}
    \sigma_{\rm rev} &=  \sum\limits_{S\in\{R,W\}}\left\{ \left(k_{S,1}^+ P_E^s-k_{S,1}^- P_{ES_1}\right)\ln\frac{k_{S,1}^+ P_E^s}{k_{S,1}^- P_{ES_1}}\right.\\
    &+\left.\sum\limits_{i=2}^n \left(k_{S,i}^+ P_{ES,i-1}^s - k_{S,i}^- P_{ES,i}^s\right)\ln\frac{k_{S,i}^+ P_{ES,i-1}^s}{k_{S,i}^- P_{ES,i}^s}\right\}\,.
    \end{split}
\end{equation}
The entropy production per product is then $\Delta\sigma = \sigma_{\rm rev}/\mathcal{J}_{\rm uni}$.

\begin{figure*}[tp]
    \centering
    \begin{subfigure}{0.32\linewidth}
        \includegraphics[width=\linewidth]{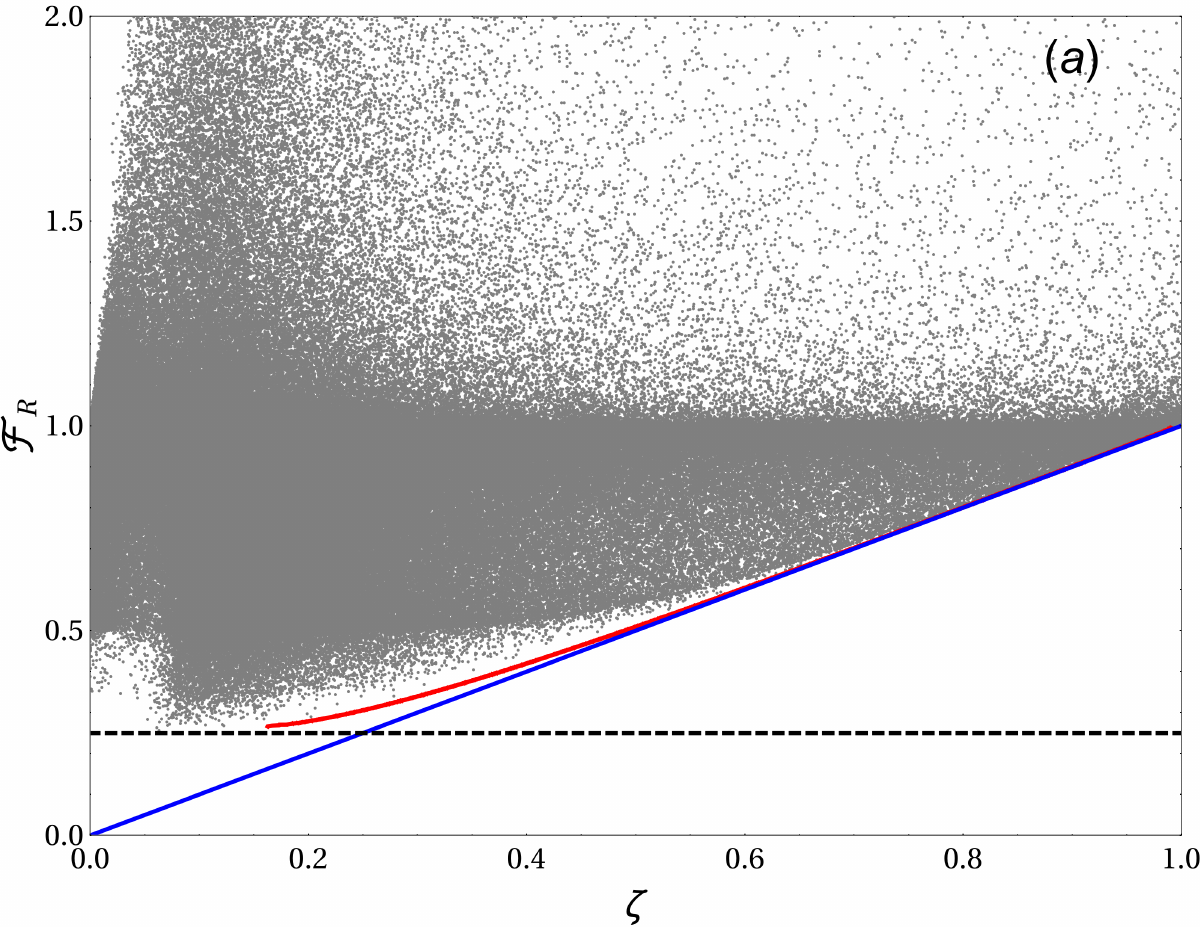}
    \end{subfigure}
    \begin{subfigure}{0.33\linewidth}
        \includegraphics[width=\linewidth]{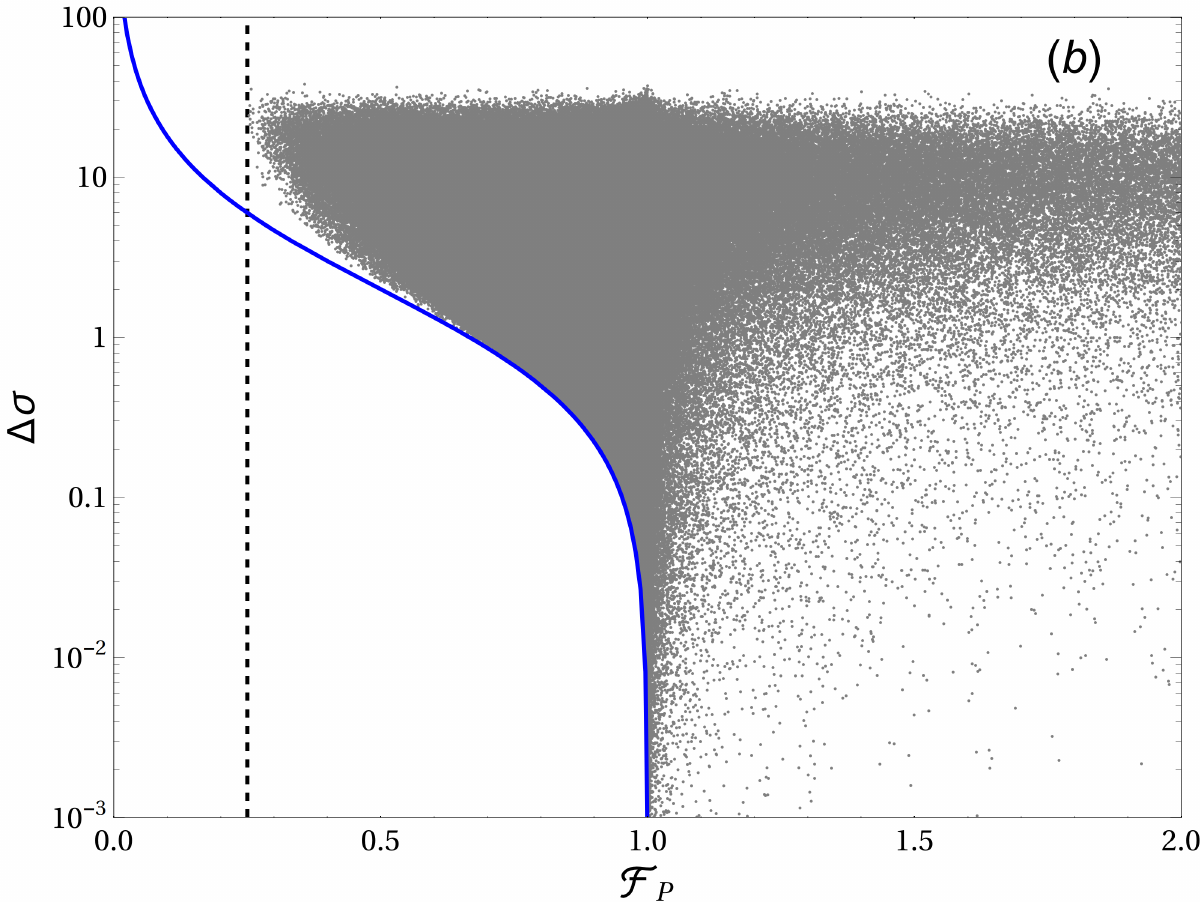}
    \end{subfigure}
    \begin{subfigure}{0.32\linewidth}
        \includegraphics[width=\linewidth]{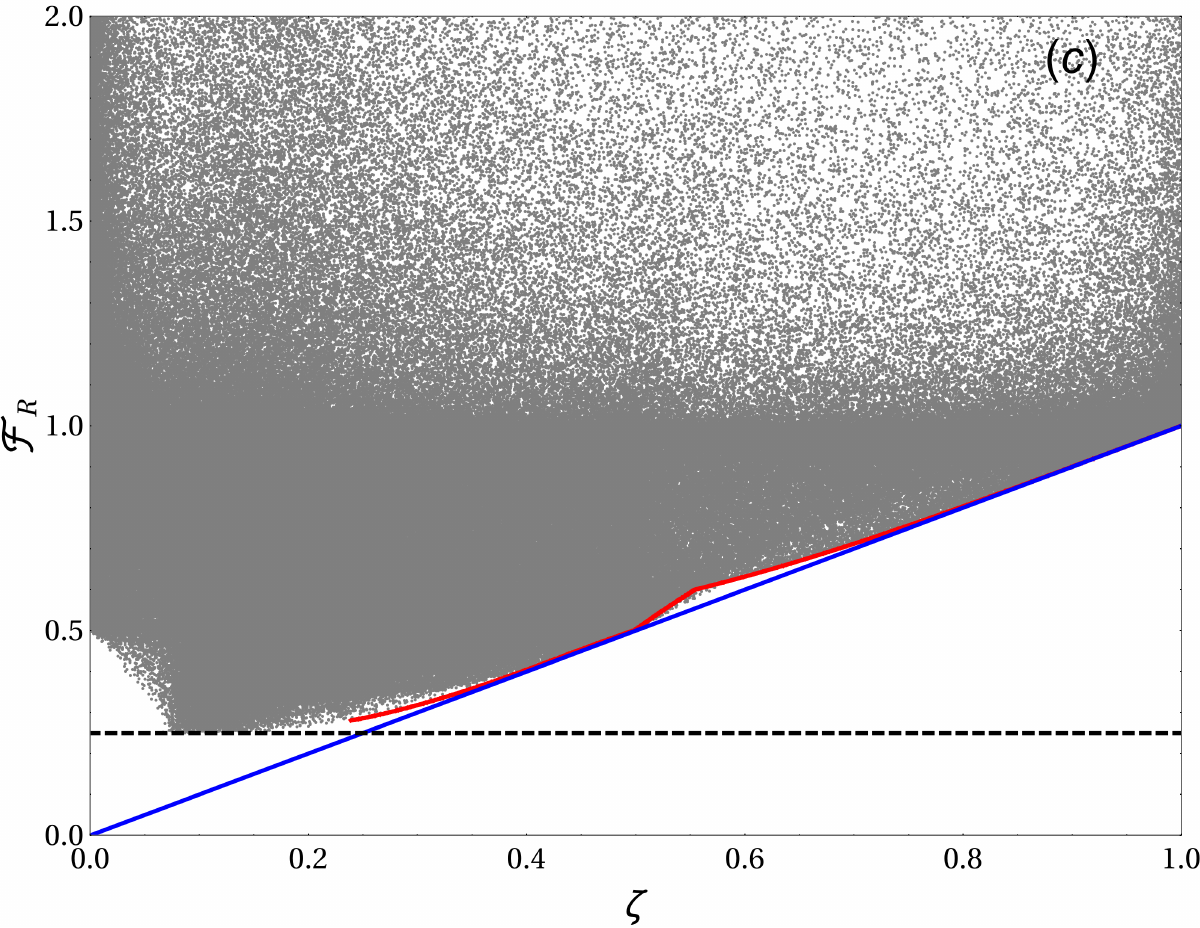}
    \end{subfigure}
    \caption{{\bf (a,c)} Lower bounds for $\mathcal{F}_R$ as a function of $\zeta = (1-\eta)/(1+\frac{\Delta\sigma}{2})$ for the $n=3$ multi-stage MM and correlated MM discrimination, respectively. {\bf (b)} Log-plot of entropy production $\Delta\sigma$ as a function of the indiscriminate Fano factor $\mathcal{F}_{P}$ for the $n=3$ multi-stage MM. Dashed black lines indicate the lower bound \eqref{eq:general_bound}, with $M =4$. Blue lines are our results \eqref{eq:TUR_Fano} for (a,c) and \eqref{eq:TUR_combined_dissipation} for (b); red lines are the Pareto optimal fronts. Gray symbols are simulated results log-uniformly generated by $10^6$ samples of the systems with random kinetic rates in the range $[10^{-3},10^3]$.}
    \label{fig:MM}
\end{figure*}

In fig.~\ref{fig:MM}(a), we show for the $n=3$ multi-stage MM scheme the lower bound on the correct product Fano factor $\mathcal{F}_R$, given by eq.~\eqref{eq:TUR_Fano}. It can be easily seen that our bound is very tight for values of $\zeta$ close to unity, while it becomes less tight for decreasing $\zeta$. However, our bound is still generally better than the well-known bound 
\begin{equation}
    \label{eq:general_bound}
    \mathcal{F}\geq \frac{1}{M}\,,
\end{equation}
used in statistical kinetics, for values of $\zeta \geq 1/M$, where $M$ is the number of states in a single production cycle \cite{David1987,Moffitt2010,Moffitt2014}. For the $n$-stage MM scheme, $M = n+1$. This bound can formally be improved upon by explicitly taking into account the thermodynamic affinity driving the production cycle \cite{Barato2015_2}, although this improvement pertains only to networks with a single affinity between substrate and product and hence does not apply to discriminatory networks. For values of $\zeta\leq 1/M$, however, the bound \eqref{eq:general_bound} is higher than our bound \eqref{eq:TUR_Fano}, so we expect our lower bound on the Fano factor of the correct products $P_R$ to have the general form
\begin{equation}
    \label{eq:TUR_bound_general_form}
    \mathcal{F}_R \geq \begin{cases}
        \frac{1}{M} & 0\leq\zeta\leq\frac{1}{M}\\
        \zeta & \frac{1}{M}\leq\zeta\leq1\,.
    \end{cases}
\end{equation}

For the Fano factor of wrong production cycles, the results are identical, but with a different $\zeta'$ that is defined as $\zeta' \equiv \eta/(1+\frac{\Delta\sigma}{2})$. Hence, in the remainder of this work, we will only consider correct product transitions. In fig.~\ref{fig:MM}(b), however, we check that the bound on the entropy production for indiscriminate product formation, i.e, eq.~\eqref{eq:TUR_combined_dissipation}, holds. It can be seen that indeed the entropy production rate per product is bounded from below by $2(\mathcal{F}_P^{-1} -1)$ for $\mathcal{F}_P<1$, and by zero otherwise.

\subsection{Michaelis-Menten network with correlations}\label{subsec:MM_corr}

We now turn to a model of MM discrimination where correlations are present between production steps \cite{Rao_2015,Poulton2019}, as can for instance occur for copolymerisation processes \cite{Johnson1993,Bauer2012}. Production in such systems is influenced not only by the addition of a new nucleotide to a growing strand but also by the nucleotide previously incorporated. This grants, \emph{e.g.,} polymerases molecular mechanisms to identify mismatches and respond appropriately \cite{Gaspard2016}.

\begin{figure}[!htp]
    \centering
    \includegraphics[width=\linewidth]{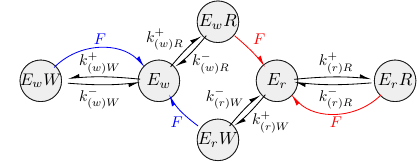}
    \caption{The MM discriminatory network with correlation between production steps. The irreversible catalysis steps are shown in red (blue) for correct (wrong) product formation.}
    \label{fig:MM_corr_network}
\end{figure}

Two consecutive steps are now not independent anymore, by assuming that a single right or wrong catalysis step affects the system discriminatory power for the subsequent substrate. We consider two parallel reaction schemes: one where a wrong product precedes the substrate binding, indicated by subscripts $w$, and the other where correct production precedes substrate binding, indicated by a subscript $r$, see fig.~\ref{fig:MM_corr_network} for an illustration of the discriminatory network. The free enzymatic states $E_r,\,E_w$ are now different due to the preceding product formation.

Since we now have four irreversible catalysis transitions, \emph{i.e.}, $\mathcal{J}_{(r)R},\,\mathcal{J}_{(r)W},\,\mathcal{J}_{(w)R},\,\mathcal{J}_{(w)W}$, the unidirectional current $\mathcal{J}_{\rm uni}$ is the total of these individual fluxes.

The error rate needs to take into account contributions from both the incorporation of wrong monomers following from a previous correct one and from ones following a previous wrong monomer, \emph{i.e.}, from the general definition in eq.~\eqref{eq:error_general}, we calculate
\begin{equation}
    \label{eq:error_MM_corr}
    \eta = \frac{\mathcal{J}_{(r)W}+\mathcal{J}_{(w)W}}{\mathcal{J}_{(r)R}+\mathcal{J}_{(r)W}+\mathcal{J}_{(w)R}+\mathcal{J}_{(w)W}}\,.
\end{equation}
The reversible entropy production now includes only the monomer binding steps, 
\begin{equation}
    \label{eq:MM_corr_entropy}
    \begin{split}
    \sigma_{\rm rev} &=  \sum_{\substack{S\in\{R,W\}\\ s\in\{r,w\}}}\left(k_{(s)S}^+\,P_{E_s}^s-k_{(s)S}^- P_{E_{s}S}\right)\ln\frac{k_{(s)S}^+P_{E_s}^s}{k_{(s)S}^- P_{E_{s}S}}\,,
    \end{split}
\end{equation}
and the entropy production per product formed is hence equal to $\Delta\sigma = \sigma_{\rm rev}/(\mathcal{J}_{(r)R}+\mathcal{J}_{(r)W}+\mathcal{J}_{(w)R}+\mathcal{J}_{(w)W})$. Plugging these expressions into eq.~\eqref{eq:TUR_Fano}, we see (cfr.~fig.~\ref{fig:MM}(c)) that once again our bound holds tightly for values of $\zeta$ close to unity, while the lower bound \eqref{eq:general_bound} takes over for smaller values. Hence, our general lower bound \eqref{eq:TUR_bound_general_form} holds.

From the Pareto front analysis and the numerical sampling of the parameter space, a deviation from the lower bound becomes apparent at around $\zeta\approx0.55$. This heralds a sudden transition in the optimal rate configuration of the network.

\subsection{Multi-stage kinetic proofreading}\label{subsec:KPR}

\begin{figure}[htp]
    \centering
    \includegraphics[width=\linewidth]{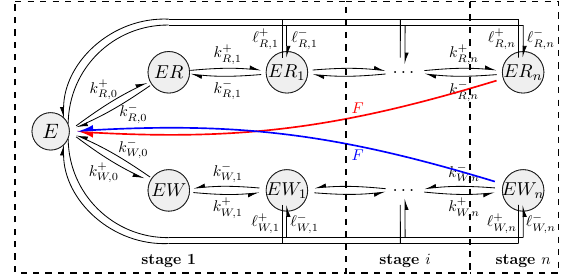}
    \caption{The multi-stage KPR discriminatory network. The irreversible catalysis steps are shown in red (blue) for correct (wrong) product formation.}
    \label{fig:multistage_KPR_network}
\end{figure}

Before moving on to our final application, we will show that our bound \eqref{eq:TUR_Fano} can be modified to include other kinetic observables. One such an observable is the cost of proofreading for the KPR model introduced by Hopfield and Ninio \cite{Hopfield1974,Ninio1975}, which is used in \emph{e.g.}, protein translation by ribosomes, or DNA polymerase. In this model, an enzyme $E$ can bind a substrate $S$ to form the complex $ES$, which can then in turn hydrolyse an energetic molecule such as ATP. This hydrolysis step brings the complex to state $ES^*$ where either product can be irreversibly formed, or the substrate can unbind, both of which restart the cycle. Multiple proofreading pathways are easily combined into a multi-stage network similar to the MM \cite{Yu2022}, with intermediate states $ES_i$, $i=1,2,...,n$ all connected to the free enzymatic state $E$, see fig.~\ref{fig:multistage_KPR_network}.

The error rate for the $n-$stage KPR model is given by the same expression as eq.~\eqref{eq:MM_error}.

Since a single cycle of the $n=1$ scheme consumes one energetic molecule, we can associate a `cost' of proofreading, which corresponds to the number of redundant cycles the system has progressed through before production \cite{Yu2022}. In short, it is the surplus energetic molecule cost per product, or, in mathematical terms
\begin{equation}
    \label{eq:proofreading_cost}
    C = \frac{\mathcal{J}_{c,R} + \mathcal{J}_{c,W}}{\mathcal{J}_R + \mathcal{J}_W}\,,
\end{equation}
with $\mathcal{J}_{c,S} = \mathcal{J}_{\ell,S}^+ - \mathcal{J}_{\ell,S}^-= \ell_{S,1}^+ P_{ES_1} - \ell_{S,1}^- P_E$ for $S\in\{R,W\}$ the proofreading costs for the correct (wrong) cycles.

Other forms and bounds can be derived from the general framework of the unidirectional TUR. In particular, bidirectional transitions can be treated as a combination of two unidirectional transitions and shifted from $\sigma_{\rm rev}$ to $\mathcal{J}_{\rm uni}$ in the definition \eqref{eq:TUR_uni_ss}, in order to generate a family of TUR bounds \cite{Pal2021_2}. If we take the proofreading transitions $ES_1\rightleftharpoons E+S$ in the single-stage KPR model to be a combination of two unidirectional transitions, the unidirectional flux $\mathcal{J}_{\rm uni}$ becomes
\begin{equation}
    \label{eq:uni_proofreading}
    \mathcal{J}_{\rm uni} = \mathcal{J}_R + \mathcal{J}_W + \mathcal{J}_{\ell,R}^+ + \mathcal{J}_{\ell,W}^+ + \mathcal{J}_{\ell,R}^- + \mathcal{J}_{\ell,W}^-\,.
\end{equation}
Rewriting now the bounds found for the Fano factors with the above assumption,  
\begin{equation}
    \label{eq:TUR_Fano_cost}
    \mathcal{F}_R \geq \frac{1-\eta}{1+\frac{\Delta\widetilde{\sigma}}{2}+C+2 r}\,,\qquad \mathcal{F}_W \geq \frac{\eta}{1+\frac{\Delta\widetilde{\sigma}}{2} +C+2r}\,,
\end{equation}
we see that the cost $C$ enters the bound, along with the normalised flux $r = (\ell_{R,1}^- + \ell_{W,1}^-) P_E^s /(\mathcal{J}_R + \mathcal{J}_W)$ from the free enzymatic state back to the bound states $ES_1$. Here, $\Delta\widetilde{\sigma}$ is the entropy production, \emph{without} taking into account the proofreading and production transitions. In this manner, suitable bounds that correspond to experimentally accessible observables can be derived to test the tightness of the TUR. Note that $r$ indicates an `upcycling' flux, where the enzyme binds a substrate and is elevated back to a high-energy conformational state, a transition which does not occur often in real proofreading systems. Hence, by assuming that $r\approx0$, we find bounds of the Fano factors \eqref{eq:TUR_Fano_cost} only involving the error rate, entropy production and proofreading cost. 

Rewriting the bound for the correct production with $r=0$, we find the following bound on the entropy production per product
\begin{equation}
    \label{eq:dissipation_rate_kpr_bound}
    \Delta\sigma \geq 2 \left(\frac{1-\eta}{\mathcal{F}_R} - (C+1)\right)\,,
\end{equation}
which is only positive when $\mathcal{F}_R \leq (1-\eta)/(1+C)$. Since a nonzero lower bound on the entropy production signifies the onset of a non-equilibrium regime, it becomes clear that increasing the proofreading cost $C$ decreases the uncertainty in the number of correct products formed, for a fixed error rate, which is at the heart of the kinetic proofreading scheme.

Continuing in this line of reasoning, let us consider what happens to the general bound on the entropy production rate \eqref{eq:TUR_combined_dissipation} when the proofreading transitions become fully irreversible, fixing $r=0$. Repeating the calculations, we find the following inequality
\begin{equation}
    \label{eq:TUR_combined_dissipation_cost}
    \Delta\sigma\geq 2 \left(\frac{1}{\mathcal{F}_P}-(1+C)\right)\,,
\end{equation}
such that a positive lower bound on the entropy production per product is found when $\mathcal{F}_P \leq 1/(C+1)$. 

For KPR, some of the kinetic rates are coupled due to the requirement that the chemical driving of the proofreading cycles is identical for both the correct and wrong pathways. For $n=1$, this amounts to the constraint
\begin{equation}
    \label{eq:KPR_constraint}
    \frac{k_{R,0}^+ k_{R,1}^+ \ell_{R,1}^+}{k_{R,0}^- k_{R,1}^- \ell_{R,1}^-} = \frac{k_{W,0}^+ k_{W,1}^+ \ell_{W,1}^+}{k_{W,0}^- k_{W,1}^- \ell_{W,1}^-}\,,
\end{equation}
which can easily be extended for the $n>1$ case.

\begin{figure*}[htp]
    \centering
    \begin{subfigure}{0.325\linewidth}
        \includegraphics[width=\linewidth]{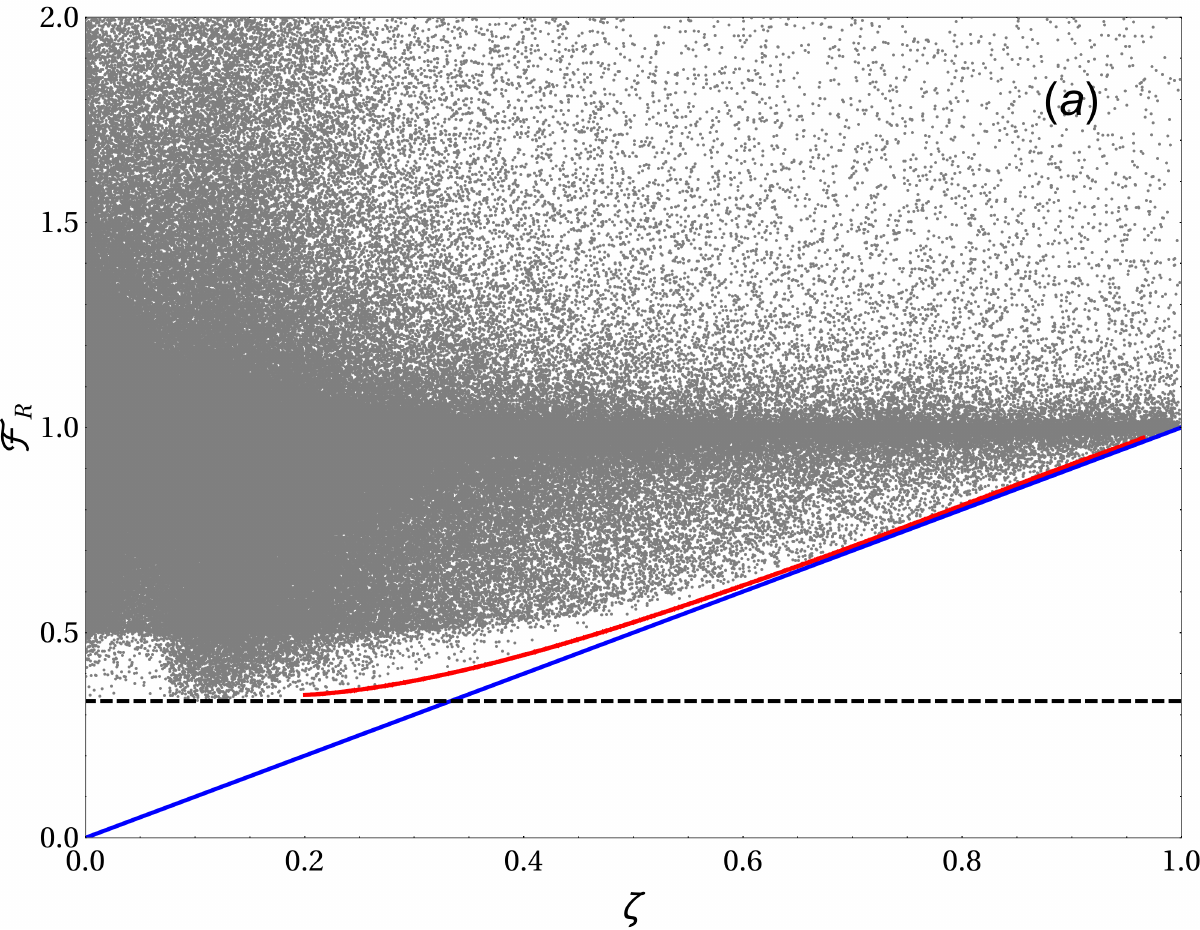}
    \end{subfigure}
    \begin{subfigure}{0.325\linewidth}
        \includegraphics[width=\linewidth]{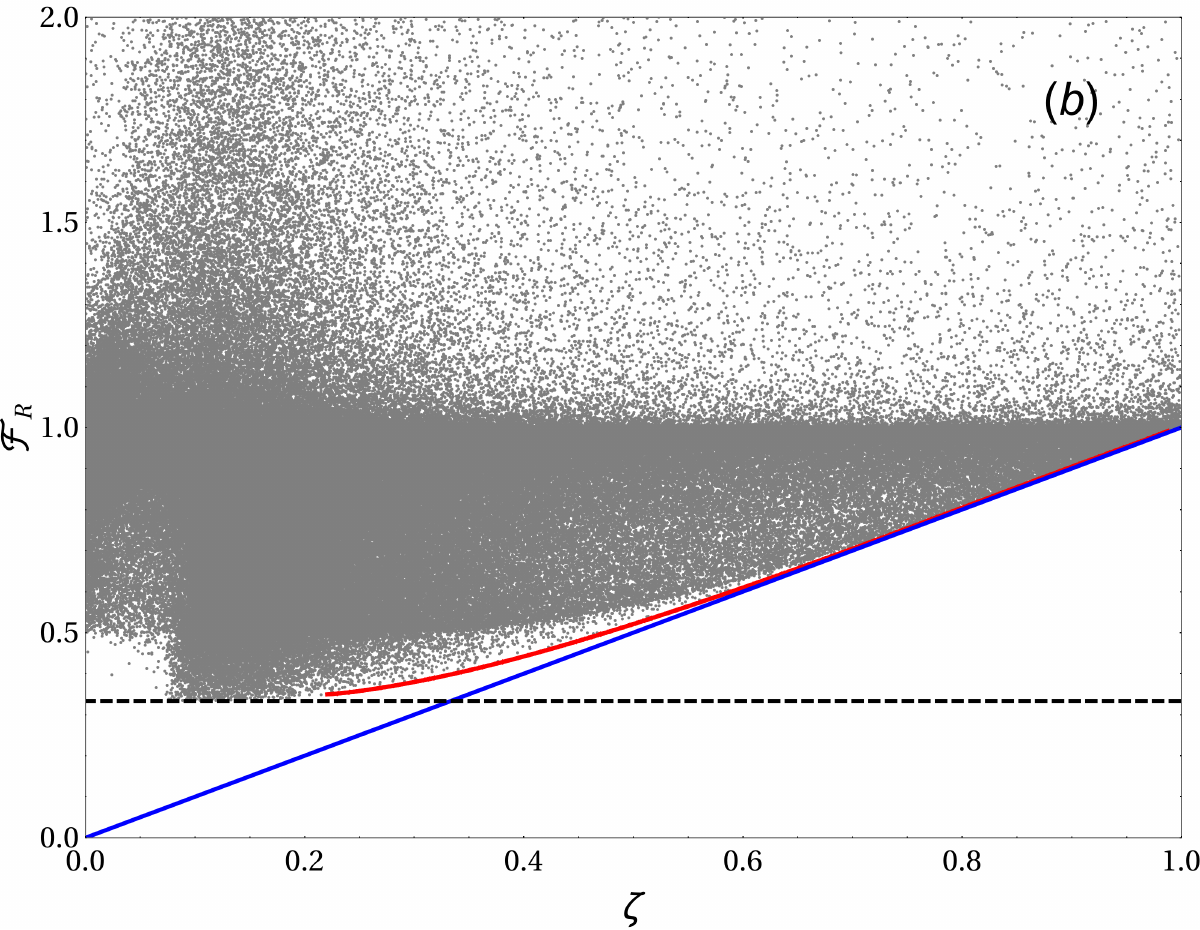}
    \end{subfigure}
    \begin{subfigure}{0.325\linewidth}
        \includegraphics[width=\linewidth]{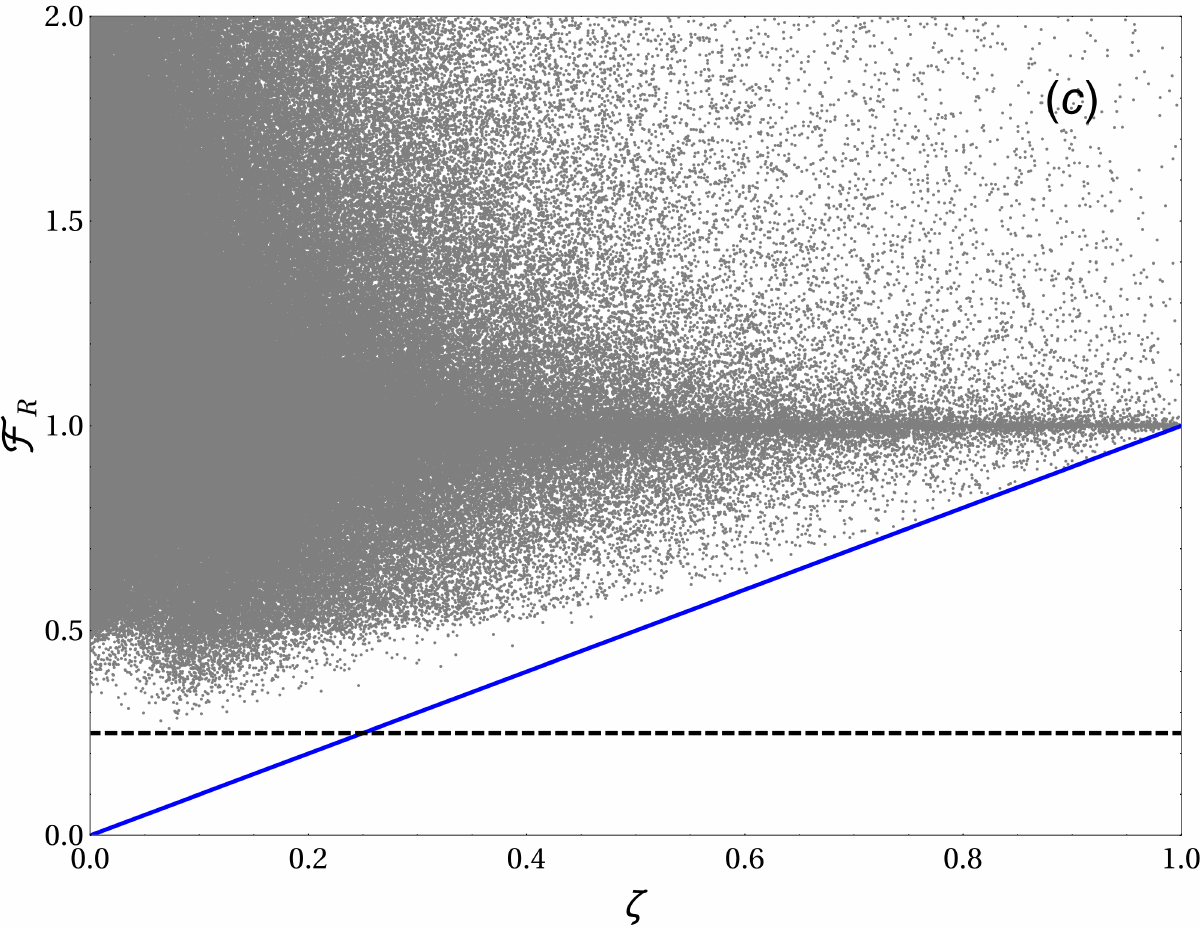}
    \end{subfigure}
    \caption{The lower bounds for $\mathcal{F}_R$ as a function of $\zeta$ for the multi-stage KPR. {\bf (a,b)} The single-stage KPR model with (a) nonzero kinetic rates, where $\zeta = (1-\eta)/(1+\frac{\Delta\sigma}{2})$, and (b) $r=0$ such that $\zeta = (1-\eta)/(1+C+\frac{\Delta\widetilde{\sigma}}{2})$. {\bf (c)} The two-stage KPR model with nonzero kinetic rates. Dashed black lines indicate the lower bound \eqref{eq:general_bound}, with $M =3$ for (a,b) and $M=4$ for (c). Blue lines are our results \eqref{eq:TUR_Fano} and red lines are the Pareto optimal fronts. Gray symbols are simulated results log-uniformly generated by $10^6$ samples of the systems with random kinetic rates in the range $[10^{-3},10^3]$.}
    \label{fig:KPR}
\end{figure*}

Our results are illustrated in fig.~\ref{fig:KPR}, where panels (a) and (c) show the Fano bound for the general $n=1$ and $n=2$ KPR scheme, respectively, with $\zeta = (1-\eta)/(1+\frac{\Delta\sigma}{2})$, and panel (b) is the alternative bound \eqref{eq:TUR_Fano_cost} with $\zeta = (1-\eta)/(1+C+\frac{\Delta\sigma}{2})$ for the single-stage KPR, by choosing the reverse kinetic rates as $\ell_W^- = \ell_R^- =0$. The lowest bound is still set by eq.~\eqref{eq:general_bound}, where now $M$ is the number of states in the cycle with largest effective length \cite{Barato2015_2}. For the multi-stage KPR, $M = n+2$. It can be seen that for all three cases our general lower bound \eqref{eq:TUR_bound_general_form} holds, for a suitable choice of the variable $\zeta$.

In fig.~\ref{fig:KPR}(c), we have not provided a Pareto optimal front, due to the computational complexity involved in calculating the Fano factor. The lower density of points close to the bound is also a consequence of the computational complexity involved in sampling the system.

\section{Conclusions \label{sec:conclusions}}
We have derived a lower bound on the Fano factors for correct and wrong products obtained through discriminatory mechanisms, as well as on the entropy production per formed product for indiscriminate production. Our analysis was based on the TUR for systems involving one or more unidirectional steps, which we assumed to be the catalysis transitions, and our results were checked against a numerically computed Pareto optimal front obtained through genetic algorithms. The bound we found is universal, independent of the details of the topology of the network, and can be applied to systems that are asymmetrical or involve more than one catalysis step per substrate. The only quantities entering our main result \eqref{eq:TUR_Fano} are the reversible entropy production per product formed and the error rate. Our bound complements the well-known statistical kinetics bound \eqref{eq:general_bound} and the bound found for systems involving only a single affinity driving the production cycle \cite{Barato2015_2}.

Moreover, we showed that the bound can be reformulated in order to include other kinetic observables, such as the hydrolytic cost in KPR. By altering our bounds, we have shown that one can alternatively obtain a lower bound on the entropy production per product formed, characterised fully by experimentally accessible quantities such as the Fano factor, the error rate or the cost.

We have illustrated the tightness of our results by analysing three discriminatory networks: i) a multi-stage MM network, a MM network with correlation between products, and a multi-stage model for kinetic proofreading.

We have restricted our analysis to the time-independent case, but it is well-known that proteins can undergo conformational transitions under the influence of time-dependent drive \cite{Stigler2011}, for which a version of the original TUR exists \cite{Koyuk2020}. In future studies, it would be interesting to study the influence of such external driving on the bounds derived here for discriminatory networks. It would also be interesting to explicitly contrast our results with the relation derived in \cite{Pineros2020}, in order to explicitly determine the influence of the catalysis transitions on the Fano bounds.

\acknowledgments
We thank Prashant Singh for carefully reading the manuscript.

J.B. is funded by the European Union’s Horizon Europe framework under the Marie Sk\l odowska-Curie grant agreement No. 101104602. K.P. is funded by the European Union’s Horizon 2020 research and innovation program under the Marie Sklodowska-Curie grant agreement No. 101064626 ’TSBC’, and from the Novo Nordisk Foundation (grant No. NNF18SA0035142 and NNF21OC0071284).

\bibliographystyle{eplbib}
\bibliography{biblio.bib}

\end{document}